%% file: GKPStabilization.tex
\documentclass[
letter,
reprint,
a4paper,
amsmath,
]{revtex4-1}

\usepackage{mathtools,amssymb,lipsum}
\usepackage{booktabs}
\usepackage{siunitx}
\usepackage[]{hyperref}
\usepackage{multirow}
\usepackage{bbold} 
\usepackage{float}
\usepackage{placeins}
\usepackage{multirow}
\input{ion_macros}
\usepackage[acronym]{glossaries}

\usepackage{color}

\let\oldcite\cite
\renewcommand{\cite}[1]{\mbox{\oldcite{#1}}}

\usepackage{tikz}
\usetikzlibrary{arrows,calc,matrix,fit,positioning}
\usetikzlibrary{decorations.pathreplacing}
\usetikzlibrary{decorations.pathmorphing}
\usetikzlibrary{decorations.markings}
\usetikzlibrary{shapes.geometric}
\begin{document}

\title{Error correction of a logical grid state qubit by dissipative pumping}

\author{B.~de Neeve}
\thanks{These authors contributed equally to this work.}
\author{T.-L.~Nguyen}
\thanks{These authors contributed equally to this work.}
\author{T.~Behrle}
\author{J. P. Home}
\email[Corresponding author, Email:]{jhome@phys.ethz.ch}

\affiliation{Institute for Quantum Electronics, ETH Z\"urich, Otto-Stern-Weg 1, 8093 Z\"urich, Switzerland}
\affiliation{Quantum Center, ETH Z{\"u}rich, 8093 Z{\"u}rich, Switzerland}

\date{\today}

\begin{abstract}
Stabilization of encoded logical qubits using quantum error correction is key to the realization of reliable quantum computers. While qubit codes require many physical systems to be controlled, oscillator codes offer the possibility to perform error correction on a single physical entity. One powerful encoding for oscillators is the grid state or GKP encoding \cite{Gottesman2000, Fluehmann2019, CampagneIbarcq2020}, which allows small displacement errors to be corrected. Here we introduce and implement a dissipative map designed for physically realistic finite GKP codes which performs  quantum error correction of a logical qubit implemented in the motion of a single trapped ion. The correction cycle involves two rounds, which correct small displacements in position and momentum respectively. Each consists of first mapping the finite GKP code stabilizer information onto an internal electronic state ancilla qubit, and then applying coherent feedback and ancilla repumping. We demonstrate the extension of logical coherence using both square and hexagonal GKP codes, achieving an increase in logical lifetime of a factor of three. The simple dissipative map used for the correction can be viewed as a type of reservoir engineering, which pumps into the highly non-classical GKP qubit manifold. These techniques open new possibilities for quantum state control and sensing alongside their application to scaling quantum computing.
\end{abstract}

\pacs{}

\maketitle

Quantum error correction is expected to be critical to scaling quantum computers to sizes which are required for practical tasks \cite{ Preskill2018}. As a result, investigations of error correction form a major part of current-day research into quantum computing \cite{IARPALogiQ}. For qubit-based approaches, in which protected logical qubits are realized in entangled states of multiple physical systems, realizing control of even minimal instance quantum error correction codes is challenging \cite{andersen2020, Nigg2014, egan2020}. The challenge is significantly reduced when considering error correction codes formed in the higher dimensional Hilbert space of a single oscillator, which has led to early demonstrations of encoding and correction with these systems \cite{ofek2016, reinhold2020, CampagneIbarcq2020, Fluehmann2019}. One of the most promising oscillator codes is the Gottesman-Kitaev-Preskill (GKP) encoding, that corrects for small displacements of the oscillator in phase space. This encompasses many of the most relevant physical errors which occur on an oscillator \cite{Gottesman2000, albert2018}. The ideal code space for the symmetric square GKP code  can be formed as the eigenspace of two stabilizer operators $S_z = e^{2 \sqrt{\pi} i q}$ and $S_x = e^{-2 \sqrt{\pi} i p}$ which are modular functions of the position and momentum variables (we define $q = (a^\dag + a)/\sqrt{2}$, $p = i(a^\dag - a)/\sqrt{2}$, $[q,p]=i$). The logical operators for this code are also displacements, with half the amplitude $Z_L = e^{\sqrt{\pi} i q}$ and $X_L = e^{-\sqrt{\pi} i p}$ ($Y_L = iX_L Z_L$). The exact eigenfunctions of these operators are periodic arrays of displaced position eigenstates, which are unphysical (infinite GKP states). However these states can be approximated by states formed as a finite energy superposition of position-squeezed states with relative weights following a Gaussian envelope - the computational basis states can for example be written as a superposition of displaced squeezed-vacuum states $\ket{r} = S(r) \ket{0} $ as $\ket{z}_L = N_z \sum_{s=-\infty}^{\infty} e^{-\pi \kappa^2 (2s+z)^2} e^{i(2s+z)\sqrt{\pi}p} \ket{r}$ with $z = 0, 1$ and  $\kappa=e^{-r}$ the reduced uncertainty in the r.m.s. position quadrature compared to the ground state. Here we use the position squeezing operator  $S(r) = e^{i\frac{r}{2}(qp + pq)}$ with $r$ real. Approximate GKP qubits were recently realized both with trapped ions \cite{Fluehmann2019} and in superconducting circuits \cite{CampagneIbarcq2020}, with the latter also demonstrating error correction through the use of modular variable measurements and conditional feedback. In both cases, logical and stabilizer measurements were limited by the fact that the finite states are not eigenstates of the measurement. This reduces the measurement fidelity and complicates the error correction cycle \cite{CampagneIbarcq2020}.

In this Letter, we demonstrate quantum error correction of a finite GKP encoded logical qubit using the motional degree of freedom of a trapped ion. We design and implement a dissipative process for which the steady state contains the logical codespace. For the feedback cycle as well as for state measurement, we introduce modified modular variable operators which are specifically designed for the finite GKP states. We show that these produce higher fidelity measurement outcomes and allow better preservation of the underlying state compared to the modular variable measurements employed in earlier work \cite{CampagneIbarcq2020, Fluehmann2019}. In combination with conditional coherent feedback and optical pumping these realize the dissipative pumping which we use to prepare and stabilize logical qubits stored in the square and hexagonal finite GKP codes. 

\begin{figure}
    \centering
    \includegraphics[width=1\textwidth]{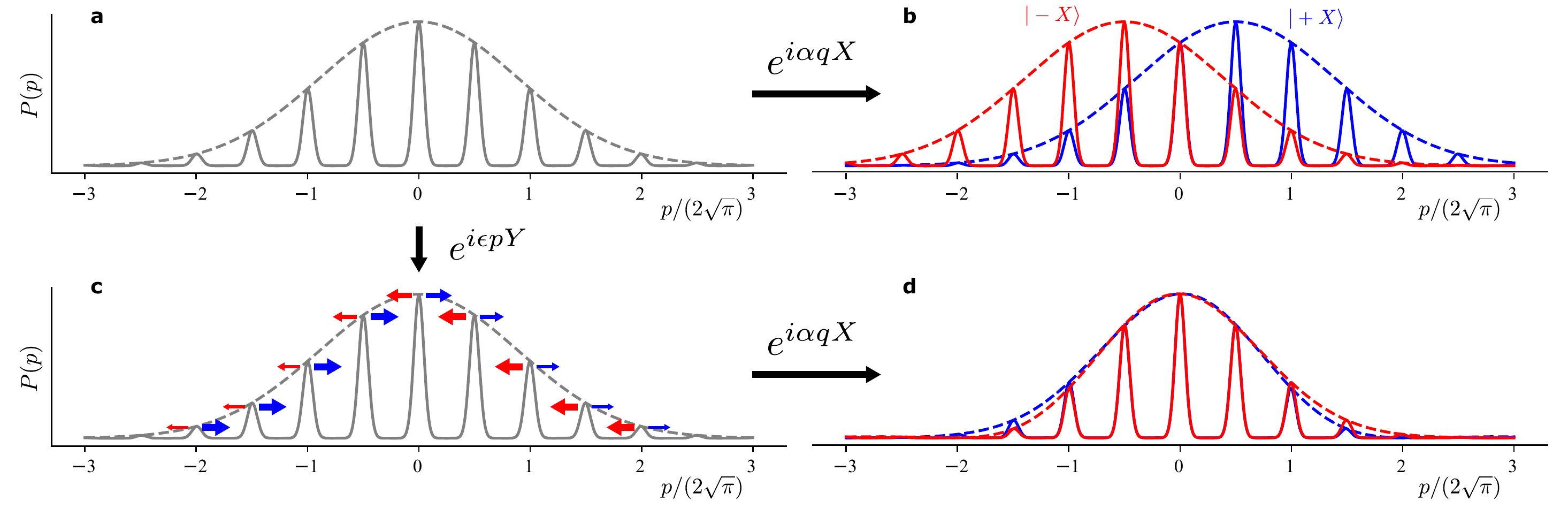}
    \caption{Comparison of the infinite state measurement to the finite state measurement. a) The probability density of a finite GKP state in the momentum space. b) The infinite measurement of position displaces the momentum distributions differently depending on the internal state $X$ eigenvalue. This imperfect overlap limits the contrast of subsequent internal state measurements in the $Y$ and $Z$ bases, which are used to extract logical and stabilizer information, and also distorts the finite state envelopes. c) The addition of a prior $e^{i \epsilon p Y}$ pulse biases the weights of the $X$ eigenstates dependent on the momentum (indicated by the thickness of the red and blue arrows). d) Under the subsequent measurement pulse, the two components then show high overlap, and the envelope is preserved. }
    \label{fig:finitetheory}
\end{figure}

The GKP encoding protects against small displacement errors in phase space, which can be detected through a modular variable measurement. This can be implemented by coupling an initial motional state $\rho_m$ to a two-level ancilla system prepared in $\ket{0}_S$  through the conditional displacement $e^{i \alpha q X}$ (here and in what follows $X, Y, Z$ are the Pauli operators for the ancilla). The outcome of a subsequent ancilla qubit measurement along $Y$ is given by $P(\pm Y) = (1 \pm \expect{Y})/2$, where the expectation value of $Y$ is $\expect{Y} = \mathrm{Tr}(M_Y \rho_m)$ with $M_Y(\alpha) = \sin(2 \alpha q)$, while for a measurement along $Z$ the relevant operator is $M_Z(\alpha) = \cos(2 \alpha q)$ \cite{Fluehmann2018}. We use $\alpha = k \sqrt{\pi}/2$, with  $k = 1$ for a logical operator and $k = 2$ for a stabilizer. As an example, when a measurement is applied to a finite GKP code state displaced along the position quadrature by a distance $\chi$ (relevant to measurement errors), the expectation value of the measurement is well approximated by $\expect{Y}_{\chi} = (-1)^{kz} e^{-\pi k^2 \kappa^2/4} \sin(k\sqrt{\pi}\chi)$ (discussions of the approximations can be found in \cite{Methods}). The pre-factor $e^{-\pi k^2 \kappa^2/4}$ is due to the finite squeezing of the code states, and results from the state-dependent displacements used in the measurement producing two displaced squeezed states which have a non-unity overlap, as illustrated in figure \ref{fig:finitetheory} b). In performing error correction, the measurement outcome for a stabilizer ($k = 2$) is used to condition the sign of a feedback displacement $e^{\pm i \mu p}$, which acts to correct the displacement $\chi$. However in the case of the finite code, the envelope of the measured state becomes increased relative to the pre-measurement state, reducing the  overlap between the pre and post-measurement states.

To mitigate these limitations, we introduce a modified modular variable measurement. The basic insight, illustrated in figure \ref{fig:finitetheory} c) and d), is to try to maintain the finite envelope in the presence of the measurement displacements. This requires that the measurement operator $e^{i \alpha q X}$ acts on an $X$ basis spin superposition which is biased depending on the momentum quadrature. To achieve this, we precede the state-dependent measurement displacement with another state-dependent displacement $e^{i \epsilon p Y}$ which rotates the spin depending on the momentum, resulting in the unitary $e^{i\alpha q X} e^{i \epsilon p Y}$ . The measurement outcome for a subsequent spin measurement along $Y$ for a displaced finite GKP state is then well approximated by \cite{Methods}
\be
\label{eq:M}
\expect{Y}_{\epsilon, \chi} = \expect{Y}_{\chi} \left[e^{-\frac{\epsilon ^2}{\kappa ^2}} + \sin \left(k \sqrt{\pi } \epsilon \right)\right] \ .
\ee
The measurement signal can thus be increased for suitably chosen $\epsilon$. 

\begin{figure}
    \centering
    \includegraphics[width=1\textwidth]{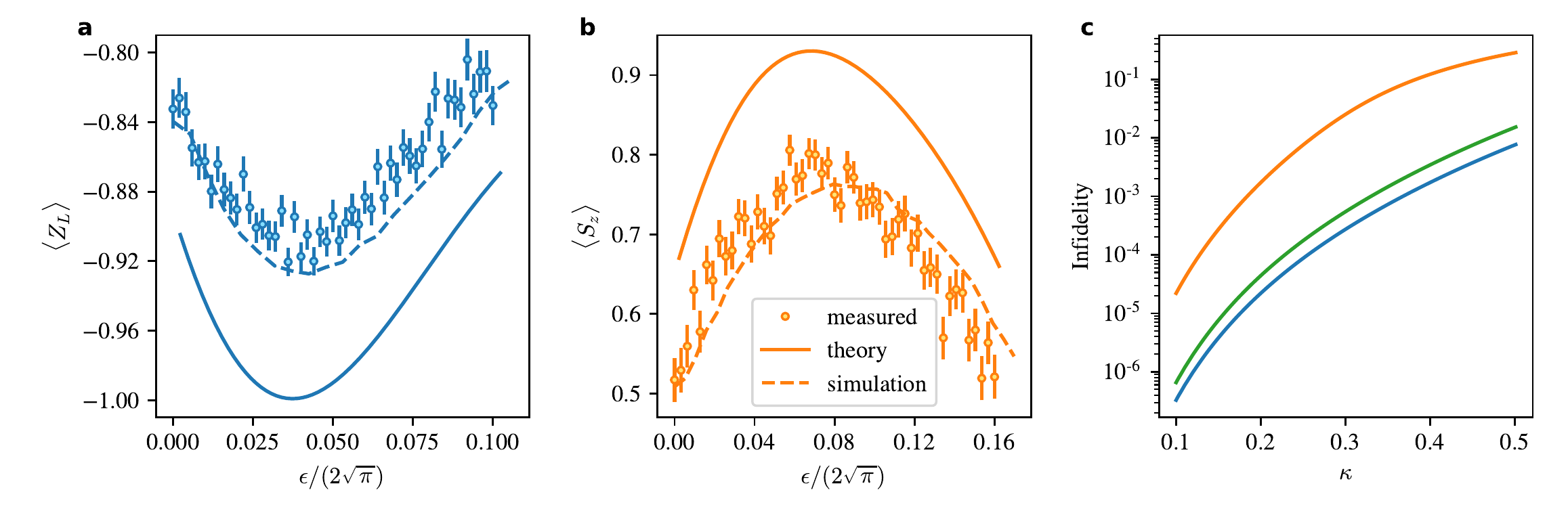}
    \caption{Finite state measurement of a) $\langle Z_L \rangle$ and b) $\langle S_z \rangle$ for a deterministically prepared GKP $\ket{1_L}$ state. Good agreement is found between the optimal value of $\epsilon$ in experimental data and equation  \ref{eq:M} (solid curve). The theoretical curves are shifted by $2\sqrt{\pi}\times0.0023$ to the right to account for an effective offset due to pulse shaping. The discrepancy in  values is consistent with known noise sources, which are accounted for in our simulations  (dashed curves, see \cite{Methods}). Error bars here and in the subsequent plots are given as  standard error on the mean. c) Infidelity of stabilizer (orange), logical (blue) measurements and preservation infidelities (green) predicted by equations \ref{eq:M} and \ref{eq:F}  as a function of $\kappa$. The experiments operate close to $\kappa=0.37$.} 
    \label{fig:finitesizedata}
\end{figure}

This insight also allows us to design a simple dissipative stabilization sequence for the finite GKP code. We use the measurement sequence above for encoding information regarding errors onto the ancilla pseudo-spin, but instead of the spin measurement, the unitary part is followed by a further state-dependent displacement operation $e^{i \mu p Y}$ which performs a coherent controlled feedback on the oscillator. Optical pumping of the spin then removes entropy. A single cycle of correction is formed from two rounds of a completely positive map $\rho_m' = \Xi_i(\rho_m) = {\rm Tr}_S\left(U_i\rho_m\otimes\ket{0}_S\bra{0}_S U_i^\dagger\right)$
with $U_1 = e^{i \mu p Y} e^{i\alpha q X} e^{i \epsilon p Y}$ and $U_2 = e^{i \mu q Y}e^{-i\alpha p X} e^{i \epsilon q Y}$ for correction in position and momentum respectively. ${\rm Tr}_S$ refers to a trace over the ancilla. For this process, we are interested that states are pumped towards a steady state and in the ideal case the fidelity of an error-free state should not be reduced by the pumping process. We choose $\epsilon$ to approximately fulfil the latter condition by maximizing the preservation fidelity $F_{\rm pres} = \bra{z_L} \rho_m' \ket{z_L}$  of the logical state after one application of $\Xi_1$ to the initial state  $\rho_m = \ket{z_L} \bra{z_L}$. We find this to be optimal for $\mu = \epsilon$ with the fidelity well approximated by
\be
F_{\rm pres}(\epsilon) = e^{-\frac{\pi \kappa^2}{2} }\left(e^{-\frac{\epsilon^2}{\kappa^2}}+\sin(\epsilon \sqrt{\pi})\right)^2 \ , \label{eq:F}
\ee
which is subtly different from equation \eqref{eq:M}. This is not analytically solvable but can be maximized for $\epsilon$ numerically. A plot of 1-$F_{\rm pres}(\epsilon_{\rm opt})$ as a function of $\kappa$ is given in figure \ref{fig:finitesizedata}. The experiments described below operate close to $\kappa = 0.37$. This two-round feedback offers a simplification over the earlier work in superconducting cavities, where additional feedback steps were introduced to counter the effect of envelope diffusion \cite{CampagneIbarcq2020}. 

The reason to apply a dissipative map rather than measurement and classically controlled feedback in the trapped ion setting is that ancilla measurement relies on  one of the two ion internal states scattering many (of order 1000) photons. Each photon scattered results in recoil displacements for absorbtion and emission of photons, leading to diffusion which would overwhelm the code. Our correction uses only internal state repumping, which involves scattering an average of $\sim 2$ photons, reducing the feedback time and producing a corresponding average displacement which we estimate to be a quadrature shift of $||(\delta_q,\delta_p)|| =0.13 $ \cite{Methods}. This is at a level which is correctable by the code. The stabilization is similar to performing pulsed sideband cooling which is commonly performed in trapped ion experiments \cite{95Monroe}.

We demonstrate the measurement and stabilization technique using the axial motional mode of a single trapped \caf ion with a frequency of around $\omega_m \approx 2 \pi \times \Uni{1.7}{MHz}$. The motional mode is controlled and read out via the internal electronic levels $\ket{0}_S \equiv \ket{^{2}{S}_{1/2}, m_j=1/2}$ and $\ket{1}_S \equiv \ket{^{2}{D}_{5/2}, m_j=3/2}$. The state-dependent displacements are implemented by application of an internal state-dependent force based on a bi-chromatic laser pulse simultaneously driving the red and blue motional sidebands of the qubit transition, realizing the Hamiltonian $\hat{H} = \hbar \eta \Omega \sigma_{\phi_s} q_{\phi_m}/\sqrt{2}$ where $q_{\phi_m} = \cos(\phi_m)q - \sin(\phi_m)p$ and $\sigma_{\phi_s} \equiv \left(\cos(\phi_s) X + \sin(\phi_s) Y \right)$. Here $\eta \simeq 0.05$ denotes the Lamb-Dicke parameter~\cite{98Wineland2}. 
Exponentiating this Hamiltonian over a relevant time period $t$ realizes a displacement  $e^{-i\gamma q_{\phi_m}\sigma_{\phi_s}}$ where $\gamma = \hbar\eta\Omega t/\sqrt{2}$, which allows the spin basis, displacement direction and displacement amplitude to be selected through the choice of $\phi_s, \phi_m, \Omega$ and $t$. 

To illustrate the improvement offered by the finite-state measurement, we apply it to a finite GKP logical state generated deterministically by creating a squeezed state with 8.9~dB of squeezing by reservoir engineering \cite{Kienzler2015, Lo2015}, followed by an appropriate sequence of four state-dependent displacements \cite{hastrup2019, Methods}. Figure \ref{fig:finitesizedata} shows the results of the measurement of both a logical operator as well as the stabilizer as a function of $\epsilon$. The optimal values of $\epsilon_{\rm opt} =2 \sqrt{\pi}\times 0.042(1)$ and $2 \sqrt{\pi}\times 0.071(3)$ respectively agree with the maximization of equation \eqref{eq:M}. These produce experimental maximum values of $\expect{Y}_{\epsilon_{\rm opt}, 0}$ which provide our best estimate of the logical and stabilizer readout which are $\langle Z_L\rangle = -0.911(3)$ and $\langle S_z\rangle = 0.780(7)$ respectively.  For comparison, a perfect implementation of both state preparation and measurement would produce  maximal values of $\langle Z_L\rangle  = -0.999$ and $\langle S_z\rangle =0.930$.

\begin{figure}
    \centering
    \includegraphics[width=.8\textwidth]{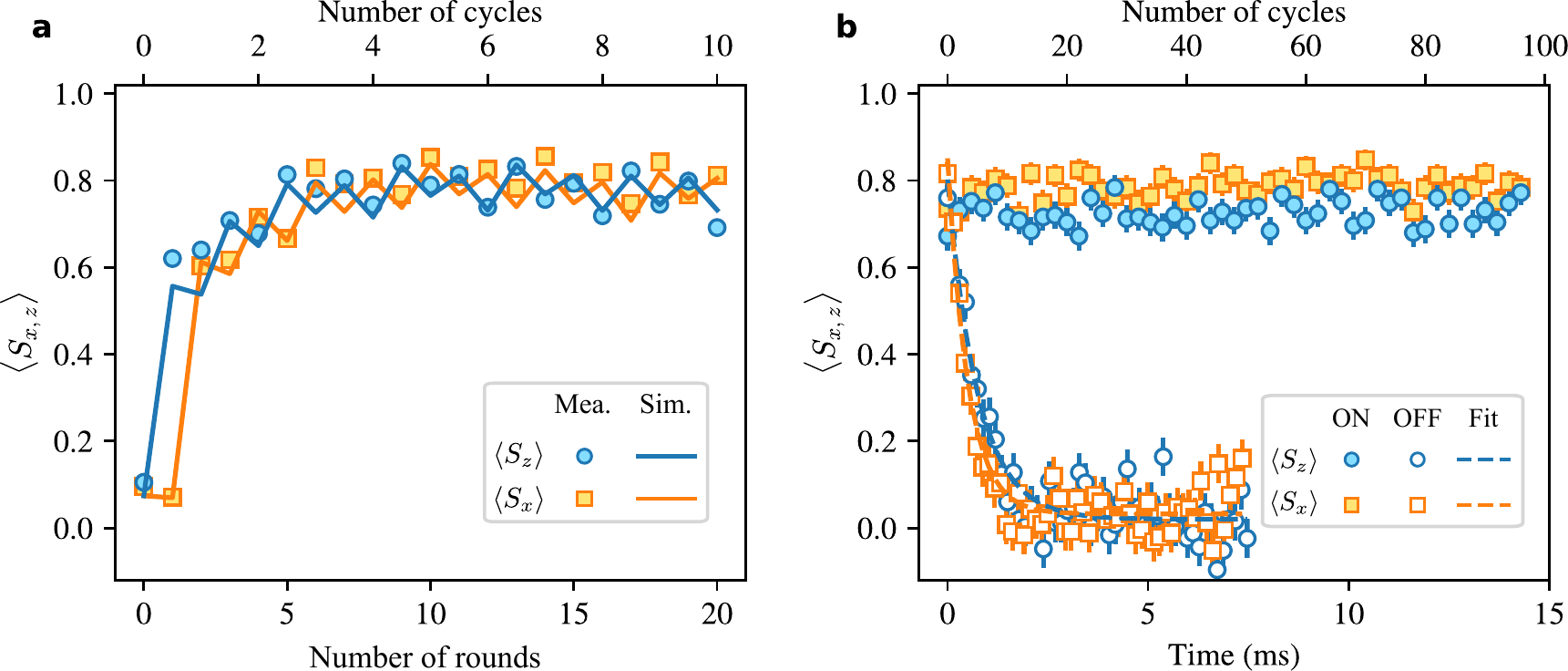}
    \caption{a) Evolution of finite stabilizer measurement values for $S_z$ and $S_x$ starting from the ground state. Steady state values are reached after 6 rounds of stabilization. b) A comparison between stabilizer readout values with (filled markers) and without error correction (open markers). An exponential fit yields lifetimes of 0.76(6) and 0.51(4)~ms for $\expect {S_z}$ and $\expect{S_x}$ respectively. }
    \label{fig:onset}
\end{figure}

We then switch our attention to the preparation and stabilization of logical qubits using the dissipative pumping. We start by initializing the ion in the ground-state using standard laser cooling \cite{95Monroe}, followed by repeated application of the error correction cycle, given by the two-round dissipative map $\Xi_2(\Xi_1(\rho_m))$ with $\epsilon \approx 2 \sqrt{\pi}\times 0.045$ and $\mu \approx 2 \sqrt{\pi}\times  0.065$ in which an effective offset of $\sim 2 \sqrt{\pi}\times  0.007$ is introduced due to pulse shaping (the value of $\mu$ was experimentally optimized for performance after a fixed number of cycles). Figure \ref{fig:onset} a) shows the evolution of both stabilizers as a function of the number of cycles. These reach a quasi steady-state of $\langle S_x \rangle$ = 0.81(3) and $\langle S_z \rangle = 0.78(4)$ after 6 cycles. This is a significant speed-up compared to the previous work which required $\sim 20$ cycles \cite{CampagneIbarcq2020}. A small  oscillation of $S_x$ and $S_z$ occurs because despite the use of a finite $\epsilon$ in a correction round, the measurement of $S_x$ slightly enlarges the envelope in $p$ and vice versa. This effect is corrected by the subsequent round. As can be seen in figure \ref{fig:onset} b), the stabilizer values persist for 94 cycles of error correction ($\sim 13$ ms) with no noticeable decay. Without applying error correction, the stabilizer values decay rapidly. An exponential fit to the non-stabilized data  yields decay time constants of $0.76(6)$~ms and $0.51(4)$~ms for $S_z$ and $S_x$ respectively.

\begin{figure}
    \centering
    \includegraphics[width=1\textwidth]{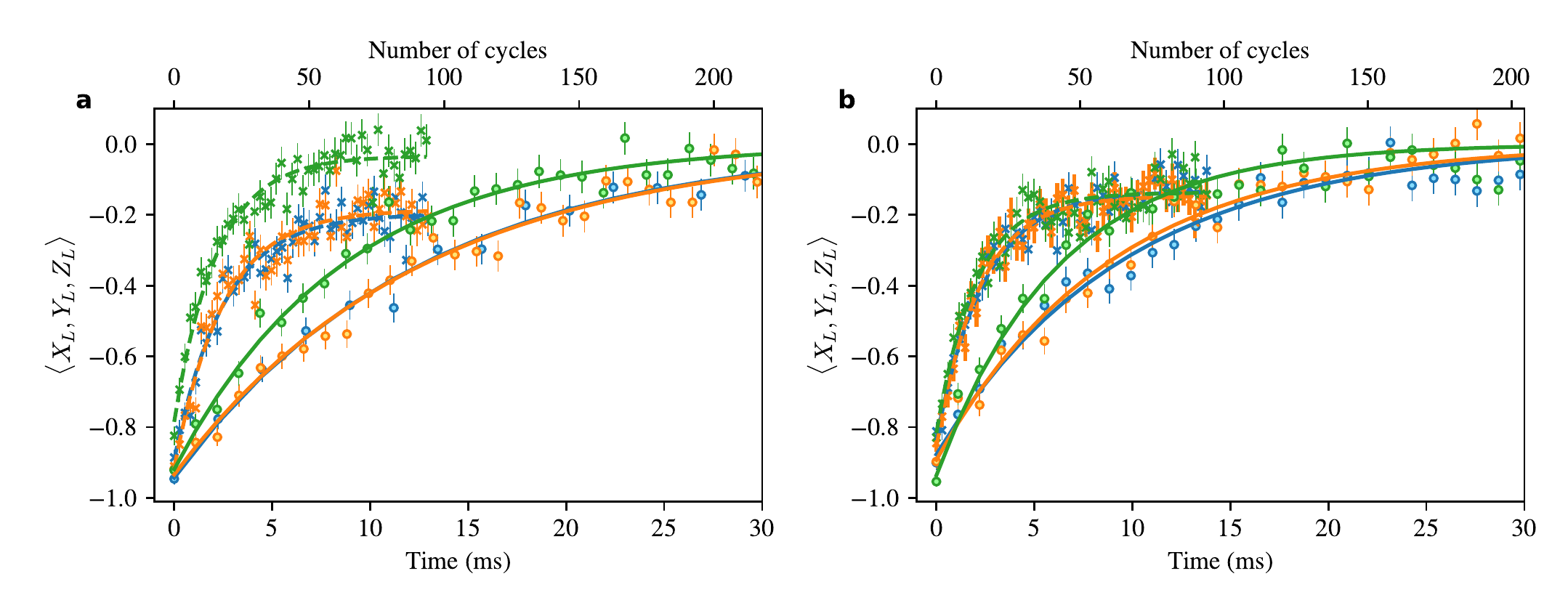}
    \caption{Logical readout of $\langle X_L\rangle$ (orange),$\langle Y_L\rangle$ (green) and $\langle Z_L\rangle$ (blue) with stabilization (circles) and without stabilization (crosses) for both the a) square and b) hexagonal finite GKP encoding. Each data set is fitted with an exponential decay. The baseline is set to zero for the case with error correction and is treated as a fit parameter for the case without error correction. a) The  logical lifetimes for the square code are $12.6(4)$, $8.6(3)$ and $12.3(5)$~ms with error correction and $2.5(2)$, $2.2(2)$ and $2.5(2)$~ms without error correction. b) For the hexagonal code, the corresponding lifetimes are $8.9(3)$, $6.2(3)$ and $9.6(4)$~ms with error correction and $2.2(1)$, $2.1(2)$ and $2.4(2)$~ms without error correction.  } 
    \label{fig:stabilization}
\end{figure}

We next compare the coherence of logical states with and without error correction. States are initialized using the same protocol as for the finite state measurement. The evolution of the logical values are plotted in figure \ref{fig:stabilization} a). For ease of viewing, data are shown for even numbers of stabilization cycles, since each cycle flips the logical state. We fit an exponential decay curve $a + b e^{-\gamma t}$ to each data set. With no stabilization, we obtain decay time constants of $(\gamma_X, \gamma_Y, \gamma_Z) = (2.5(2), 2.2(2), 2.5(2))$~ms, while when applying error correction we obtain $(12.6(4), 8.6(3), 12.3(5))$~ms. The shortest coherence time of stabilized logical states is $3.4$ times longer than the best value found without stabilization. This marks a clear improvement for all logical operators. By comparison to simulations involving independently characterized noise, we see that trap frequency noise is the primary limitation to both the unstabilized as well as the stabilized coherence times. This causes the unstabilized logical values to decay to a non-zero value because dephasing preserves the state radius and thus maintains some memory of the initial state (for more discussion of this effect and the error modelling, see \cite{Methods}). Diffusive noise such as heating will eventually bring the unstabilized logical readout to zero at a longer time scale than our measurement. The logical readout does decay to zero when applying error correction. This is expected, since stabilization keeps the oscillator state in the code space where a fully mixed logical state has a zero expectation value.  Since $Y_L$ involves a larger displacement than $Z_L$ and $X_L$, it is most affected by the finite extent of the state, resulting in a reduced readout value. Similarly, the eigenstates of $Y_L$ also decay more rapidly than the other eigenstates, since they are more strongly affected by the dominant error channels.

In an attempt to suppress the asymmetry in performance of $X_L, Y_L, Z_L$ and their eigenstates, we also implemented stabilization of the fully symmetric hexagonal GKP code \cite{CampagneIbarcq2020},  which is constructed from the stabilizer operators of amplitude $\alpha_H=\sqrt{\pi} \sqrt{2/\sqrt{3}}$ making an angle of $\phi_x = 2 \pi/3$ with each other: $S_x = e^{i\alpha_H ( \cos(\phi_x)q - \sin(\phi_x) p)}$ and $S_z = e^{i \alpha_H q}$, with logical operators $X_L = e^{i \alpha_H ( \cos(\phi_x)q - \sin(\phi_x) p)/2}$ and $Z_L = e^{i \alpha_H q/2}$. 
The displacement amplitudes from the origin for $X_L, Y_L, Z_L$ are symmetric in this code: $Y_L=i X_L Z_L = e^{i \alpha_H ( \cos(\phi_y)q - \sin(\phi_y) p)/2}$, where $\phi_y = \pi/3$.
Results for stabilization of logical states of the hexagonal code are shown in figure \ref{fig:stabilization} b), with the stabilization cycle performed using two correction rounds for $S_z$ and $S_x$ respectively. Fitting exponential decays as before, we obtain decay time constants of $(2.2(1), 2.1(2), 2.4(2))$~ms for the unstabilized vs. $(8.9(3), 6.2(3), 9.6(4))$~ms for the stabilized data. We see that the coherence is lower than for the square code, while the asymmetry remains (some asymmetry is expected from the stabilization). 


Finally, we implement a similar dissipative method for initializing the $X_L,Y_L,$ and $Z_L$ logical states from a partly mixed code state. We do this with the map $\Xi_i(\rho_m) = {\rm Tr}_S\left(U_i\rho_m\otimes\ket{0}_S\bra{0}_S U_i^\dagger\right)$. Taking $\ket {+X_L}$ as an example, the unitary is $U_{+X} = e^{i \delta(p - q) Y}e^{-i\delta p X}e^{i\delta q}e^{i \epsilon qY}$ with $\delta=\sqrt\pi/2$ (unitaries for initialization of other logical eigenstates are given in \cite{Methods}). We test this method experimentally by preparing logical states from a mixed state formed by applying 6 cycles of stabilization to the ground state of the oscillator. We subsequently measure finite state readouts of $X_L, Y_L$ and $Z_L$ are $0.91(1)(-0.91(1)), 0.83(1)(-0.84(1)), 0.82(1)(-0.86(1))$ for the corresponding $+(-)$ eigenstates prepared in this way.

The results show not only the ability to stabilize finite logical qubits for quantum computing, but are also an example of the use of reservoir engineering to produce highly non-classical oscillator states. A reduction in technical noise could allow the logical coherence to be extended beyond the coherence time of $\sim 16$~ms we obtain from a $(\ket{0} + \ket{1})/\sqrt{2}$ Fock state superposition of the oscillator (we already exceed the 1.7~ms coherence time of the spin). 
The GKP code we use may not be the best fit for our dephasing noise model, although we think that once sequences of gates are performed it may turn out to be favourable \cite{albert2018}.  The simplicity of the implementation makes it attractive for future embeddings in larger architectures.  While here we demonstrate the utility of short sequences of state-dependent displacements followed by optical pumping, an open question for future research is how extended sequences, possibly including additional unitary steps or involving multiple oscillators, could provide stabilization of a wider range of novel states and subspaces. As with any form of oscillator reservoir engineering in trapped ions our results are connected to laser cooling - in parallel work, we have shown that the modular variable measurement with autonomous feedback represents an efficient form of laser cooling with regards to the number of repumping cycles required \cite{deNeeve2020}. Finally, the states involved in these experiments are squeezed in multiple dimensions, which also makes them amenable for quantum enhanced sensing \cite{17Duivenvoorden}. These various aspects point to the applicability of our methods in a number of areas of interest for future research.

JH devised the pumping scheme and performed initial simulations, which were then extended by TLN. BN programmed the experimental sequences and experiments were carried out by TLN, BN and TB. TLN performed data analysis. The paper was written by JH with input from all authors.

We thank P. Campagne-Ibarcq for stimulating discussions. We acknowledge support from the Swiss National Science Foundation through the National Centre of Competence in Research for Quantum Science and Technology (QSIT) grant 51NF40–160591, and from the Swiss National Science Foundation under grant number 200020 165555/1. JH thanks E.~M.~Home, P.~D.~Home and Y.~Iida for allowing time for the theoretical part of this work by distracting the children.

While we were performing experiments and writing this paper, we became aware of parallel theoretical work on finite GKP state measurement \cite{hastrup2020} and on finite GKP state stabilization \cite{royer2020}.


\bibliography{./myrefs2}

\FloatBarrier

\onecolumngrid

\clearpage
{\LARGE\bfseries Supplementary Information\\}

\section{Theoretical background}
\subsection{Kraus operators}
An important component of any of the calculations given in the main text are the Kraus operators. Two processes occur with relevant Kraus operators. The first is the finite state measurement, while the second is the map involved in the feedback cycle. For measurement, using an ancilla prepared in $\ket{0}$ and a unitary operation $e^{i\alpha q X} e^{i \epsilon p Y}$ as given in the main text, we can choose to project the ancilla in the $Z$ or $Y$ bases. We find the Kraus operators for the $Z$ ancilla projection
\be
K_{\pm}^Z = \left(e^{i \alpha q} \cos(\epsilon p - \pi/4) \pm  e^{-i \alpha q} \cos(\epsilon p + \pi/4)\right)/\sqrt{2} 
\ee
while for a  $Y$ projection we obtain
\be
K_{\pm}^Y = \left(\cos(\alpha q) e^{\pm i \epsilon p} - \sin(\alpha q) e^{\mp i \epsilon p} \right)/\sqrt{2}.
\ee

The measurement operator is given by
\be
M = K^\dagger_+ K_+ - K^\dagger_- K_-
\ee
with the respective $K_\pm$. This results in a list of 16 terms, all of which are products of displacement operators.

For the stabilization sequence, the $K_{\pm}^Y$ operators are modified by the feedback controlled-displacement. We then find 
\be
K_{\pm}^F =  \left(e^{i\mu p }\cos(\alpha q) e^{\pm i \epsilon p} \pm e^{-i \mu p} \sin(\alpha q) e^{\mp i \epsilon p}\right)/\sqrt{2}
\ee
These can also be arranged into products of the form $e^{i A q} e^{i B p} e^{i\phi_{AB}}$.

\subsection{Finite GKP state calculations}
Equations \ref{eq:M} and \ref{eq:F} are evaluated using the Kraus operators given above. This is simplified if the correct choice is made for the representation of the finite GKP states. The major displacement in any measurement is by $\alpha = k \sqrt{\pi}/2$ with $k = 1,2$. This performs displacements along $p$.  If the GKP state is considered in the $q$ representation, it consists of a sum of separated squeezed states with only minimal overlap between neighbours. Furthermore under the displacement $e^{i \pm \alpha q}$ neighboring squeezed states do not interfere. Since the displacements $e^{\pm i \epsilon p}$ and $e^{\pm i \mu p}$ are smaller than the spacing between squeezed states, we can also make the approximation that these do not produce significant overlap between neighboring squeezed states, which greatly simplifies calculations. 

The $\ket{0_L}$ finite GKP logical state can be written as
\be
\ket{0_L} = N_0 \sum_{s = -\infty}^{\infty} e^{-\kappa^2 (2 s l)^2/2} e^{2 i s l  p} S(r) \ket{0},
\ee
where $e^{2 i s l  p}$ is a translation along the position axis $x$ by $2 s l p$ and $\ket{0}$ is the harmonic oscillator ground state. The squeezing operator reduces the extent of the state along the $q$ axis, following $S(-r)q S(r) = q \kappa$, where $e^r = 1/\kappa$. For the symmetric square GKP state, $l = \sqrt{\pi}$. The normalization of the state is given by
\be
1 = N_0^2 \sum_{s,s' = -\infty}^{\infty} e^{-\kappa^2 ((2 s l)^2 + (2 s' l)^2)/2}  \bra{0} e^{-2 i s l  p/\kappa}  e^{2 i s' l  p/\kappa} \ket{0} .
\ee
Assuming that neighboring squeezed states do not have significant overlap, then
\be
N_0^2 = \left( \sum_{s = -\infty}^{\infty} e^{-\kappa^2 (2 s l)^2} \right)^{-1} \ .
\ee

The measurement operator is given by a sum of displacements with phases, thus our problem reduces to terms evaluated on the ground state of motion which are all of the form
\be
\bra{0} e^{-2 i s l  p/\kappa} e^{i A q \kappa} e^{i B p/\kappa}  e^{i\phi_{AB}} e^{2 i s l  p/\kappa} \ket{0}
\ee
where $A$ and $B$ are real-valued displacement amplitudes and $\phi_{AB}$ is a scalar phase factor which arises from re-ordering operators. This evaluation can be performed using standard quantum optics, in particular the expectation value of the displacement operator on the ground state gives
\be
\bra{0} e^{i A q + i B p} \ket{0} = e^{-|A + i B|^2/4} \,
\ee
which is valid for real $A$ and $B$.

The full evaluation for any particular logical state then involves a double sum over these terms
\be
N_0^2 \sum_{s,s' = -\infty}^{\infty} e^{-\kappa^2 (2 s l)^2/2} e^{-\kappa^2 (2 s' l)^2/2} \bra{0} e^{-2 i s' l  p} e^{i A q \kappa} e^{i B p/\kappa}  e^{i\phi_{AB}} e^{2 i s l  p} \ket{0}
\ee
where $A$ and $B$ are relevant displacement values obtained from the Kraus or measurement operators.

Assuming again that initially separated squeezed states do not overlap, the  dominant terms result from $s=s'$, and we can ignore the others. Thus the relevant terms in the evaluation of the operator become
\be
N_0^2 \sum_{s = -\infty}^{\infty} e^{-\kappa^2 (2 s l)^2}  \bra{0} e^{-2 i s l  p/\kappa} e^{i A q \kappa} e^{i B p/\kappa}  e^{i\phi_{AB}} e^{2 i s l  p/\kappa} \ket{0}
\ee

For measurements requiring measurement displacements $\alpha$ along the position axis, we note that a similar procedure can be followed, but writing the finite GKP state as a sum of squeezed states narrowed in the $p$ quadrature.

\subsection{Preservation fidelity}
The fidelity in equation \ref{eq:F} compares the overlap of the pre-stabilization finite GKP state with a single round of stabilization. The action of the stabilization round on the input state $\rho_m = \ket{0_L} \bra{0_L}$ gives rise to a density matrix  
\be
\rho_m' = \Xi(\rho_m) =  K_+^F \ket{0_L} \bra{0_L} (K_+^F)^\dagger + K_-^F \ket{0_L} \bra{0_L} (K_-^F)^\dagger  \ .
\ee
Using the definition of the fidelity as 
\be
F(\rho, \ket{0_L}) = \bra{0_L} \rho \ket{0_L}
\ee
gives equation \ref{eq:F}. A similar procedure can be carried out for the other logical state.

\section{Experimental methods}
\subsection{Experimental sequence}

We summarize the sequence used for the experiments described in the main text by a series of sub-sequences, which we
group conceptually into state preparation and GKP stabilization.

\subsubsection{State preparation}

The state preparation is further divided into cooling, squeezed-state pumping, and GKP preparation. We first apply Doppler cooling, followed by EIT cooling to bring all three motional modes of a single trapped \caf ion close to their quantum mechanical ground state. We then focus our attention on the axial mode which we cool further using sideband cooling, and then prepare a squeezed state by switching the interaction to the squeezed Fock basis \cite{Kienzler2015}. Starting from a squeezed state with 8.9 dB of squeezing we then apply a sequence of four state-dependent displacements in order to generate a 4-component GKP state with an approximately Gaussian envelope \cite{hastrup2019}. Taking as an example the preparation of $\ket{1_L}$ in the square GKP code: starting from an initial position-squeezed state $\ket{r} = S(r)\ket{0}$, then the first and third state-dependent displacements are applied to split the state into a superposition of four components separated in position by $\alpha_{\rm grid} \equiv 2\sqrt{\pi}$. The second pulse is used to tune the amplitudes of the components in order to approximate a Gaussian envelope. The fourth pulse approximately disentangle the pseudo-spin from the motion. In this case we have $\ket{1_L}\otimes \ket{0}_S \approx e^{-i\alpha_4 q Y} e^{i\alpha_3 p X} e^{i\alpha_2 q Y} e^{-i\alpha_1 p X} \ket{r} \otimes \ket{0}_S$, with $\alpha_1 = \alpha_{\rm grid}, \alpha_2 = 0.031\alpha_{\rm grid}, \alpha_3 = 0.5\alpha_{\rm grid}, \alpha_4 = 0.125\alpha_{\rm grid}$. By rotating the motional phase of the squeezing and state-dependent displacements by $\frac{\pi}{2}$, this instead prepares $\ket{-X_L}$. More generally, we can write the preparation sequence using the unitary operator \[U_{\rm prep}(\theta, \lambda) = e^{i \beta_4 q_{\phi_4} Y} e^{i \beta_3 q_{\phi_3} X} e^{i \beta_2 q_{\phi_2} Y} e^{i \beta_1 q_{\phi_1} X}\] with $\phi_1 = \frac{\pi}{2} + \theta, \phi_2 = \theta, \phi_3 = -\frac{\pi}{2} + \theta, \phi_4 = \pi + \theta$, and $\beta_1 = \frac{\alpha_1}{\lambda}, \beta_2 = \lambda \alpha_2, \beta_3 = \frac{\alpha_3}{\lambda}, \beta_4 = \lambda \alpha_4$ with $\alpha_j, j=1,\hdots, 4$ as given for the square code state above. This is applied to a squeezed state appropriately aligned with the unitary interaction to prepare the desired logical state:

\begin{align}
\ket{\psi_L}\otimes\ket{0}_S \approx U_{\rm prep}(\theta, \lambda) \ket{r e^{-i2\theta}}\otimes\ket{0}_S\, , \label{eq:meas_free_prep}
\end{align}

\noindent where $\ket{r e^{-i2\theta}} = S(r e^{-i2\theta})\ket{0}$, $S(r e^{-i2\theta}) = e^{i\frac{r}{2}(q_\theta p_\theta + p_\theta q_\theta)}$, $q_\theta = \cos(\theta)q - \sin(\theta)p$, $p_\theta = q_{\theta-\frac{\pi}{2}}$. With $\theta = 0$, $\lambda = 1$ we obtain the state preparation of $\ket{1_L}$ described above. Table \ref{tab:prep_params} summarises the parameters we use to prepare $-1$ eigenstates of the square and hexagonal GKP codes. Though the state-dependent displacements are chosen to minimise the entanglement between the pseudo-spin and the motion at the end of the GKP state preparation, we additionally follow the four pulses by a detection of the internal state in order to fully disentangle the two systems. Since a bright detection scatters many photons we conditionally continue the sequence only when the detection is dark. However, unlike previous work where grid preparation relies on post-selection of several measurements where the probability of a dark detection is \( \frac{1}{2} \), here we perform only one measurement where we can make the probability of a dark detection close to 1. In particular, the rate of success depends on the amount of entanglement remaining before the detection, which in our case is $\sim 3 \%$. Thus, the measurement ensures that the state is disentangled without the cost of any photon recoils (in contrast to a repump), and with a probability of success close to 97 \%.

\begin{table}[]
\centering
\begingroup
\renewcommand{\arraystretch}{2}
\begin{tabular}{c|c|c|c|c|c|c|}
\cline{2-7}
&\multicolumn{3}{|c|}{Square code} &\multicolumn{3}{|c|}{Hexagonal code} \\ \hline
\multicolumn{1}{|c|}{Eigenstate} & $\ket{-X_L}$ & $\ket{-Y_L}$ &$\ket{-Z_L}$ &$\ket{-X_L}$ & $\ket{-Y_L}$ &$\ket{-Z_L}$ \\ \hline
\multicolumn{1}{|c|}{$\theta$} & $\frac{\pi}{2}$ &$\frac{\pi}{4}$ &0 &$\frac{2\pi}{3}$ &$\frac{\pi}{3}$ &0 \\ \hline
\multicolumn{1}{|c|}{$\lambda$} &1 &$\sqrt{2}$ &1 &$\sqrt{\frac{2}{\sqrt{3}}}$ &$\sqrt{\frac{2}{\sqrt{3}}}$ &$\sqrt{\frac{2}{\sqrt{3}}}$ \\ \hline
\end{tabular}
\endgroup
\caption{Parameters used to prepare an approximate GKP eigenstate $\ket{\psi_L}$ in equation \eqref{eq:meas_free_prep} using the measurement free method from \cite{hastrup2019}. In the notation of the main text $\ket{-Z_L} = \ket{1_L}$.}
\label{tab:prep_params}
\end{table}

\subsubsection{GKP stabilization and measurement - duration}

The durations of the various steps in the stabilization and measurement sequence are limited by laser power and the need to avoid off-resonant excitation. For the latter the state-dependent displacement pulses are smoothly turned on/off over a timescale of one microsecond. The powers are chosen such that the shortest pulse in the sequence is about $3 \mu s$ long, and thus an entire round is performed in \(\sim 75 \mu s\), leading to a stabilization cycle of \(\sim 150 \mu s\) for the square code. Optical pumping with light at 854, 866, and 397 nm takes a duration of \(\sim 10 \mu s\). 


\subsection{Logical state initialization}
\begin{figure}
    \centering
    \includegraphics[width=.9\textwidth]{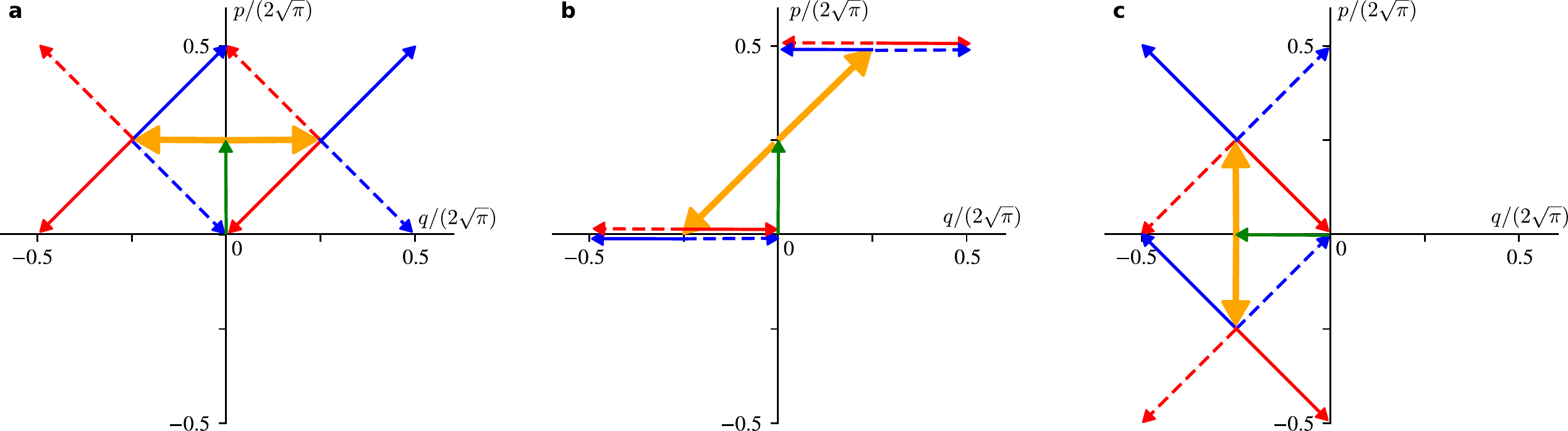}
    \caption{Illustration of displacements in phase space during logical state initialization into an eigenstate of a) $X_L$, b) $Y_L$ or c) $Z_L$. Starting at the origin, we apply a global displacement (green arrows) followed by a measurement of the logical state (thick orange arrows, short bias displacements are not shown) and an appropriate feedback (blue arrows). The wave packet is split, following different feedback paths depending on whether it is a $+$ eigenstate (red arrows) or $-$ eigenstate (blue arrows) of the logical operator. We can initialize into either $+$ or $-$ eigenstates by tuning the feedback direction as shown by the solid or dashed arrows respectively.}
    \label{SI:reset}
\end{figure}

The main idea to initialize the GKP state into an eigenstate of a logical operator is that we use a pair of state-dependent displacements to perform the corresponding finite logical measurement on an ancilla and based on this measurement outcome apply an appropriate feedback bringing to the desired GKP state by another state-dependent displacement.  In our experiments, the ancilla which is the ion's internal electronic state is prepared in $\ket 0$. Depending on whether the GKP state is an $+$ or $-$ eigenstate of the logical operator, the logical measurement rotates the ancilla by $2\pi$ or $\pi$ around the $X$ axis. Inserting a global displacement in between the two pulses of the logical measurement further rotates the ancilla an angle equal to the geometric phase acquired by the three displacements. Choosing this additional phase to be $\pi/2$, the ancilla is in $\ket{+Y}$ or $\ket{-Y}$ for $+$ or $-$ logical eigenstate. This allows us to apply a corresponding feedback displacement conditioning on the ancilla state in the $Y$ basis. Detailed pulse sequence is given in table \ref{tab:reset_params} for each logical eigenstate. The lengths and directions of each displacement on the $q, p$ phase space are illustrated in figure \ref{SI:reset}. We start with the short state-dependent displacement to assure that the subsequent finite logical measurement is correctly biased. In general, the GKP state is enlarged and not centered at the origin at the end of this sequence. We apply two extra cycles of stabilization to correct for the envelop. We observed no significant improvement in the fidelity of state preparation by repeating this procedure multiple times.

\begin{table}[]
\centering
\begingroup
\renewcommand{\arraystretch}{2}
\begin{tabular}{|c|c|}
\hline
{Eigenstate} & Sequence \\ \hline
$\ket{+X_L}$ &$e^{-i(\delta q - \delta p)Y} e^{-i\delta p X} e^{i\delta q} e^{-i\epsilon q Y}$\\ \hline
$|-X_L\rangle$ &$e^{+i(\delta q - \delta p)Y} e^{-i\delta p X} e^{i\delta q} e^{-i\epsilon q Y}$\\ \hline
$|+Y_L\rangle$ & $e^{-i\delta p Y} e^{i(\delta q-\delta p) X} e^{i\delta q} e^{i(\epsilon q+\epsilon p) Y} $      \\ \hline
$|-Y_L\rangle$ & $   e^{+i\delta p Y} e^{i(\delta q -\delta p) X} e^{i\delta q} e^{i(\epsilon q + \epsilon p) Y} $       \\ \hline
$|+Z_L\rangle$ &$    e^{-i(\delta q + \delta p)Y} e^{i\delta q X} e^{i\delta p} e^{i\epsilon p Y} $       \\ \hline
$|-Z_L\rangle$ &  $  e^{+i(\delta q + \delta p)Y} e^{i\delta q X} e^{i\delta p} e^{i\epsilon p Y} $       \\ \hline

\end{tabular}
\endgroup
\caption{Pulse sequence used for projective preparation of a GKP eigenstate $\ket{\psi_L}$. It consists of three state-dependent displacements and one global displacement. Here $\delta = \sqrt{\pi}/2$ and $\epsilon \approx 0.06\sqrt{\pi}$. In the notation of the main text $\ket{+Z_L} = \ket{0_L}$ and  $\ket{-Z_L} = \ket{1_L}$.}
\label{tab:reset_params}
\end{table}
\subsection{Calibration}
The experiments rely on precise calibrations of various control parameters. In this section we discuss the most relevant ones, namely the trap frequency and strengths of the global and state-dependent displacements. While a state-dependent displacement is implemented using a bichromatic laser pulse driving simultaneously the red and blue sidebands of the atomic transition (see main text), a global displacement is realized by application of an oscillating voltage resonant with the trap frequency to an electrode (henceforth we refer to this as ``tickling''). The Hamiltonian in the rotating frame of the oscillator reads $H_E = eE(a e^{-i\varphi_m} + a^\dagger e^{i\varphi_m})$, where $E$ and $\varphi_m$ are the amplitude and relative phase of the electric field acting on the ion. The action of this resonant tickle is to displace the oscillator in phase space by a distance proportional to the pulse duration and with the direction controlled by $\varphi_m$.

\subsubsection{Trap frequency calibration}
To calibrate the trap frequency, we use the displacements generated by the tickle. We first cool the ion to the ground state, and then apply the tickle in two sequential pulses of the same amplitude and duration $\tau$. The second pulse has opposite phase to the first. If the pulses are on resonance with the ion, the second pulse will return the ion to the ground state.  If the pulses are not on resonance then the state finishes at a displacement  $d\propto \tau^2\delta$ away from the phase space origin. We probe the final state of the ion with a red sideband probe which cannot be driven if the ion is in the ground state. This allows us to achieve \Uni{50}{Hz} precision with a pulse sequence of $< \SI{250}{\micro\second}$.  

\subsubsection{State-dependent displacement strength}
The state-dependent displacement allows us to read out the characteristic function of the motional state \cite{Fluehmann2020}. We use this as a calibration technique for the strength of a state-dependent displacement pulse by applying it to a the Fock state $\ket{n=1}$, which has a distinct feature which is not correlated with the imperfection due to finite thermal occupancy. The precision can be further improved using two orthogonal state-dependent displacements of the same length $e^{i\alpha q Y} e^{-i\alpha p X}$ on an ion  pre-prepared in $\ket{0}_S$ and cooled to a  squeezed vacuum state, squeezed along the $q$ axis (a typical value used in our experimments is $\sim 8.9$ dB of squeezing). For $\alpha = \sqrt{\pi}/2$ the internal state is maximally disentangled from the motion \cite{hastrup2019}, and the ion ends up in the bright state for the detection. Maximizing this signal provides a calibration of the displacement amplitude.

\subsubsection{Global displacement}
We calibrate the strength of a global displacement generated by the electrode tickle relative to a state-dependent displacement. We repeat the experiment used to calibrate the trap frequency, but replacing the first displacement by a state-independent displacement performed by the laser. This is implemented by preparing the ion in the superposition $(\ket{1}_S + \ket{1}_S)/\sqrt{2}$, followed by the state-dependent displacement $e^{i\alpha pX}$ , producing no spin-motion entanglement. This is followed by a tickling pulse, which reverses the action of the first displacement if the two pulses have opposite phase and the same magnitude of displacement. To probe this, we apply this sequence to an ion prepared in the ground state of motion, and subsequently use a red sideband pulse to probe whether the ion has returned to this state after the two displacements.

\subsection{Minimization of photon recoils}
Photon recoil occurs when repumping the ion from the excited state. A judicious choice of polarization minimizes the number of scattered photons, which helps the performance of the stabilization. Two Zeeman sub-levels of the internal electronic structure of a \caf ion are used as an ancilla qubit: $\ket{0}_S \equiv \ket{^{2}{S}_{1/2}, m_j=1/2}$ and $\ket{1}_S \equiv \ket{^{2}{D}_{5/2}, m_j=3/2}$. The reset of the ancilla at the end of each correction round is implemented using optical pumping. A \si{854}{nm} laser couples the $\ket{1}_S$ state to the short lived ${}^2P_{3/2}$ manifold which primarily decays to $^2S_{1/2}$ by emission of a \si{393}{nm} photon. The ion can decay with a small probability to the $^2D_{3/2}$, which can be repumped by an \si{866}{nm} laser and a $\sigma^+$-polarized \si{397}{nm} laser to pump the ion into the $\ket{0}_S$ state. The repump beams make a 45 degree angle to the motional mode which we use for the code, and propagate perpendicular to the magnetic field used to defined the quantization axis. To minimize the photons scattered, we use primarily  $\pi$-polarized light to couple $\ket{1}_S$ most strongly to the $^2P_{3/2}, m_j=+3/2$ state, which subsequently decays directly to $\ket{0}_S$. In this direct process only 2 photons are scattered per repump event. However there is a non-zero probability for the $^2P_{3/2}, m_j=+3/2$ state to decay back to $^2D_{5/2}$ manifold. Ensuring that these decays are also repumped to the ground state requires circularly polarized light. We thus use light which is linearly polarized with the direction of polarization being around 24 degrees from the quantization axis. This angle is found experimentally by maximizing the probability of repumping into $\ket{0}_S$ state when the \si{397}{nm} $\sigma^+$  beam is off. 

We have measured the effect of photon recoils on stabilizer readout. After GKP state preparation, we excite the ion into $\ket{1}_S$ followed by repumping, and subsequently read out the stabilizer (infinite state readout). The value for $\expect {S_x}$ reduces from 0.68(1) to 0.58(1) due to the photon recoils.

\subsection{Numerical simulation}
This section describes numerical simulations which allow us to understand the relevant error channels in the experiment. 

\subsubsection{Photon recoils}
We use a classical Monte-Carlo method to model photon recoils, following the ion internal state during a repump process using a rate-equation approach. At each step, we update the ion internal state based on the probability of making a transition. If a transition takes place, the corresponding recoil kick is calculated taking into account the beam geometry and dipole emission pattern. We then map the recoil kick into a displacement in phase space. From $\ket{1}_S$ this step is repeated until the ion reaches the $\ket{0}_S$ state. To keep the problem simple, we assume that the step size in the algorithm is short enough such that the ion makes zero or one transition at each step. We also neglect correlation between photon absorption and emission between sequential steps. On average, the ion scatters two photons and the oscillator is displaced by $e^{i(\delta_p q-\delta_q p)}$, with $\sqrt{\delta_q^2 +\delta_p^2}=0.13$.

\subsubsection{Error channels}
Two principal noise sources limit our experiments, which can be characterized in reference experiments. By measuring heating from the quantum ground state \cite{00Turchette} we obtain a heating rate of about 10 quanta/s for the motional mode. This can be modelled by two Lindblad superoperators $\sqrt{\gamma_1}\Lin{a}{\rho_m}$ and $\sqrt{\gamma_2}\Lin{a^\dagger}{\rho_m}$ with $\gamma_1\approx \gamma_2= \Uni{10}{s}^{-1}$. A larger impediment to current experiments is frequency fluctuations of the trap. One component of this comes from mains noise, which we can measure by performing measurements of the trap frequency as a function of delay from the line cycle. Data of this type is found in figure \ref{SI:noise}. We fit this using a superposition of multiple tones of \Uni{50}{Hz} , giving amplitudes of $(25, 1, 29, 3,31)$~Hz for the $(50, 100, 150, 200, 250)$~Hz components, which produces a functional form $\delta(t)$. In models we insert this by sampling from a time-dependent term onto the Hamiltonian $\hbar\delta(t)a^\dagger a/2$. The sampling phase is fixed or randomly sampled by the simulation to emulate experiments with or without line-trigger. In addition to line noise, we also observe a slow drift of the trap frequency $\hbar\delta_0a^\dagger a$ over the time taken to acquire data. We model this by randomly sampling a Gaussian distribution of width $2\pi\times\Uni{6}{Hz}$. Finally we find a better agreement between data and theory over a range of experiments (including those described in earlier work \cite{17Kienzler}) by introducing small component of Markovian dephasing jump operator $a^\dagger a$ at rate $\gamma_3= 20\ {\rm s}^{-1}$. 
 
\subsubsection{Monte-Carlo simulation}
\begin{figure}
    \centering
    \includegraphics[width=.9\textwidth]{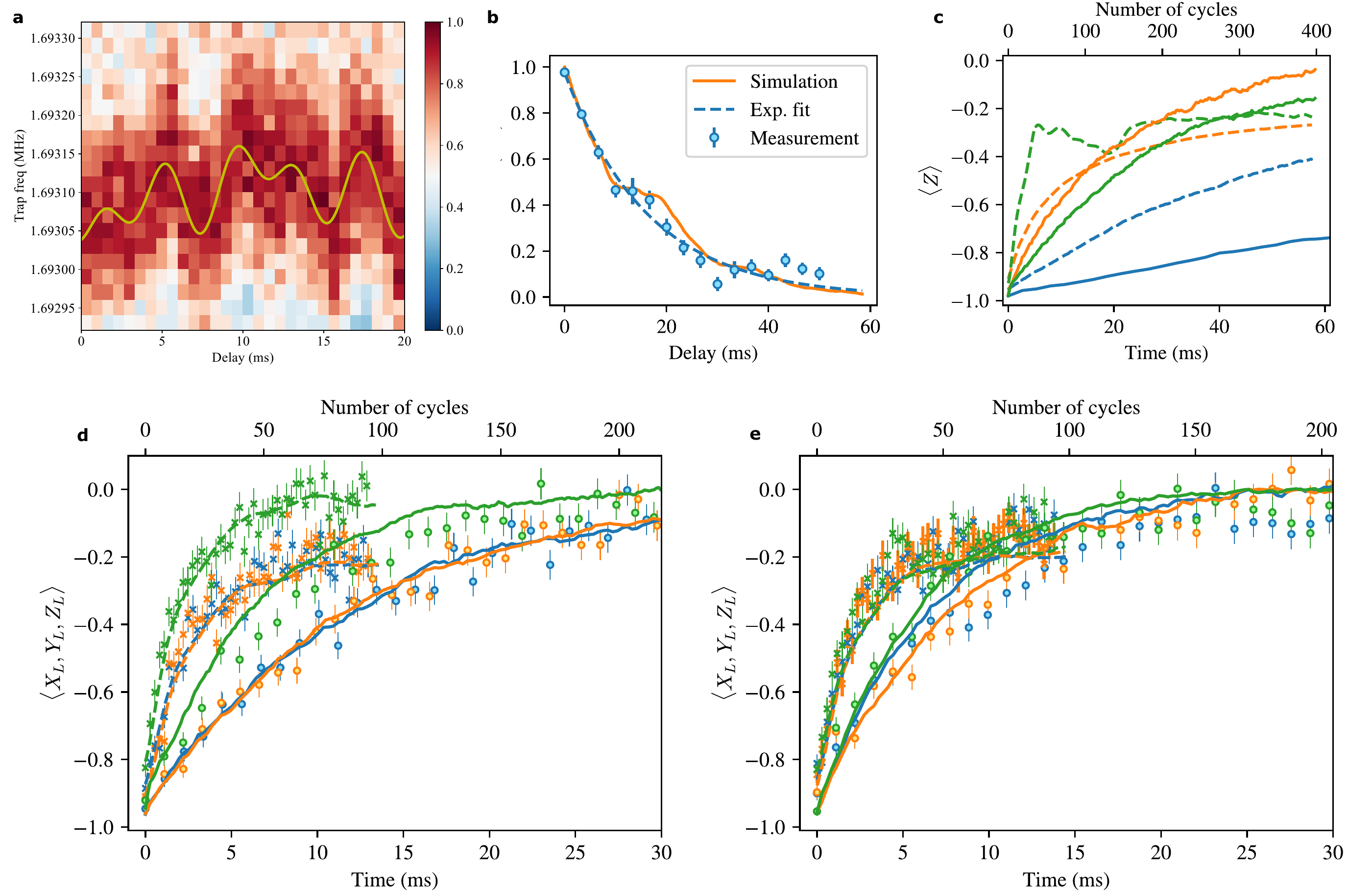}
    \caption{a) Trap frequency measurement measured at a time delayed from a 50 Hz line trigger. The trap frequency is measured using the modified tickling method. The color shows the probability of finding the ion in $\ket{0}_S$. The solid line is a fit using the five lowest harmonics of 50 Hz. b) A measurement of motional coherence of the state $(\ket{0} + \ket{1})/\sqrt{2}$, taken using a Ramsey sequence. An exponential fit yields a coherence time of 16.4(9) ms. Solid line is the Monte-Carlo wavefunction simulation based on the noise parameters given in the text. c) Decay simulation of a GKP $\ket{1_L}$ logical state under the independent action of each error channel: Markovian dephasing (orange), 50 Hz noise (green) and heating (blue) with (solid lines) and without (dashed lines) stabilization. For each error channel we optimize the frequency at which we apply the error correction. d-e) Comparison of data with simulation for the time evolution of logical readouts $\expect{X_L}$ (orange), $\expect{Y_L}$ (green) and $\expect{Z_L}$ (blue) with stabilization (solid lines) and without stabilization (dashed lines) for both the d) square and e) hexagonal finite-GKP encoding. The simulation reproduces qualitatively the experimental data for both. } 
    \label{SI:noise}
\end{figure}

To model the experiments, we use a  Monte-Carlo wavefunction approach, which allows us to sample from non-Markovian noise sources as well as to include photon recoils. Since the experiments consist of blocks of state-dependent displacements, the simulation is built based on fundamental blocks calculating the time evolution of the oscillator--ancilla system. The mains noise and the static frequency offset $\delta_0$ are randomly drawn and fed to the simulation at the beginning of each trajectory. At the end of each error correction round, we determine whether the ion is in $\ket{1}_S$ state, which then selects a photon recoil displacement. We use a modified version of the Monte-Carlo solver from QuTiP \cite{qutip1, qutip2}, which allows us to efficiently parallelize the calculation. Results are shown in figure \ref{SI:noise} for square and hexagonal code using the noise parameters extracted from comparison experiments, with the addition of a small amount of Markovian dephasing.
\subsection{Additional data}
\begin{figure}
    \centering
    \includegraphics[width=.9\textwidth]{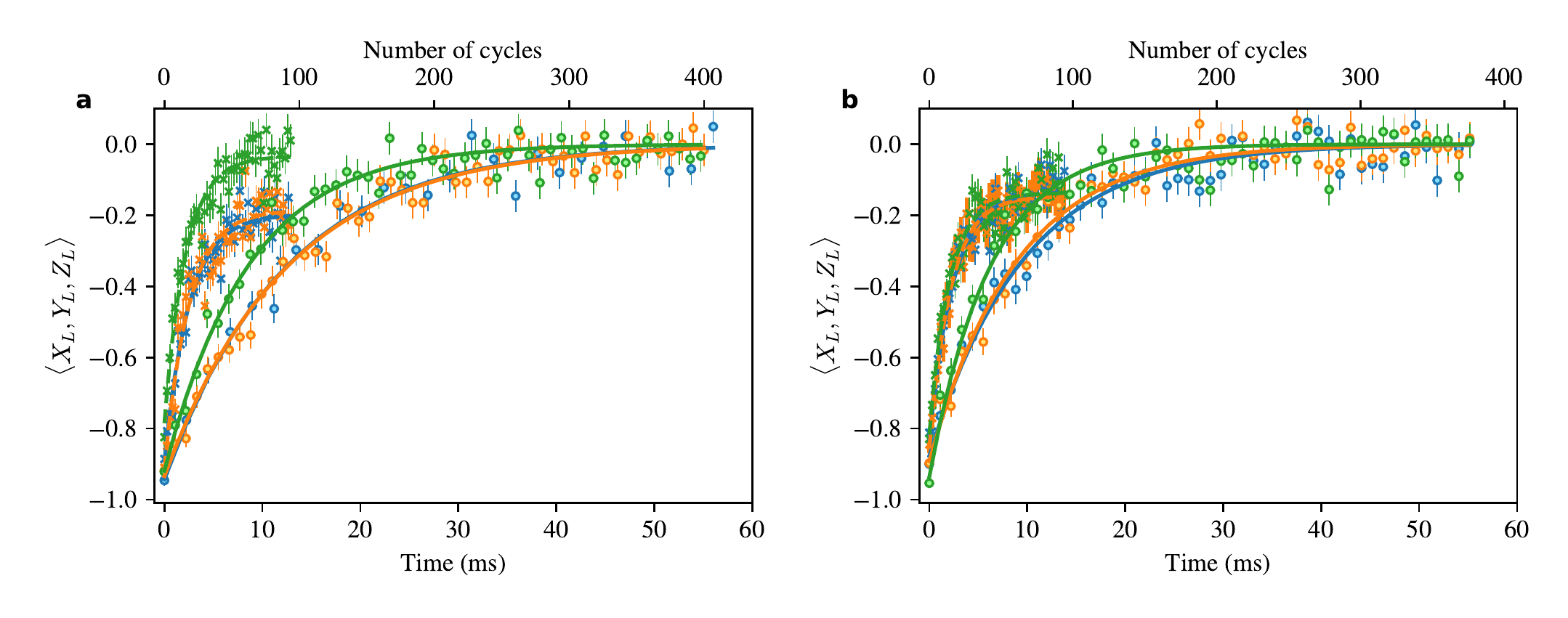}
    \caption{Full datasets for logical readouts of $\expect{X_L}$ (orange), $\expect{Y_L}$ (green) and $\expect{Z_L}$ (blue) with stabilization (circles) and without stabilization (crosses) for both a) square and b) hexagonal finite-GKP encoding. Exponential fits are also shown for the cases with stabilization (solid lines) and without stabilization (dashed lines). The short-time parts of both data sets are shown in the main paper in figure \ref{fig:stabilization}.} 
    \label{SI:fulldata}
\end{figure}

Figure \ref{SI:fulldata} shows full data taken for logical readouts of $\expect{X_L}$, $\expect{Y_L}$ and $\expect{Z_L}$ with stabilization and without stabilization for both square and hexagonal finite GKP encoding. The stabilization introduces additional diffusion which eventually brings the logical values to zero.

\end{document}

%% file: ion_macros.tex
\newcommand{\Uni}[2]{\ensuremath{{#1} \mathrm{\:{#2}} }}
\newcommand{\Lin}[2]{\mathcal{D}[#1](#2)}

\newcommand{\ket}[1]{\ensuremath{\left| {#1} \right>} }
\newcommand{\bra}[1]{\ensuremath{\left< {#1} \right|} }
\newcommand{\expect}[1]{\ensuremath{\langle #1 \rangle}}

\newcommand{\be}{\begin{eqnarray}}
\newcommand{\ee}{\end{eqnarray}}

\newcommand{\caf}{\ensuremath{^{40}{\textrm{Ca}}^{+}\, }}